\documentclass[%
 reprint,
footinbib,
 amsmath,amssymb,
 aps,
prl,
]{revtex4-1}

\usepackage{graphicx}
\usepackage{dcolumn}
\usepackage{bm}

\begin{document}

\title{A Unified Picture of Lattice Instabilities in Metallic Transition Metal Dichalcogenides}
\author{Diego Pasquier}
\email{diego.pasquier@epfl.ch}
\affiliation{Institute of Physics, Ecole Polytechnique F\'{e}d\'{e}rale de Lausanne (EPFL), CH-1015 Lausanne, Switzerland}
\author{Oleg V. Yazyev}
\email{oleg.yazyev@epfl.ch}
\affiliation{Institute of Physics, Ecole Polytechnique F\'{e}d\'{e}rale de Lausanne (EPFL), CH-1015 Lausanne, Switzerland}

\begin{abstract}
Transition metal dichalcogenides (TMDs) in the $1T$ polymorph are subject to a rich variety of periodic lattice distortions, often referred to as charge density waves (CDW) when not too strong.
We study from first principles the fermiology and phonon dispersion of three representative single-layer transition metal disulfides with different occupation of the $t_{2g}$ subshell: TaS$_2$ ($t_{2g}^1$), WS$_2$ ($t_{2g}^2$), and ReS$_2$ ($t_{2g}^3$) across a broad range of doping levels.
While strong electron-phonon interactions are at the heart of these instabilities, we argue that away from half-filling of the $t_{2g}$ subshell, the doping dependence of the calculated CDW wave vector can be explained from simple fermiology arguments, so that a weak-coupling nesting picture is a useful starting point for understanding. 
On the other hand, when the $t_{2g}$ subshell is closer to half-filling, we show that nesting is irrelevant, while a real-space strong-coupling picture of bonding Wannier functions is more appropriate and simple bond-counting arguments apply. Our study thus provides a unifying picture of lattice distortions in $1T$ TMDs that bridges the two regimes, 
while the crossover between these regimes can be attained by tuning the filling of the $t_{2g}$ orbitals. 
\end{abstract}

\maketitle
Layered transition metal dichalcogenides (TMDs) have been the subject of much attention, to a large extent due to the occurrence of a rich variety of lattice instabilities \cite{wilson_charge-density_1974, wilson_charge-density_1975, whangbo1992analogies, castro_neto_charge_2001, rossnagel2011origin, manzeli_2d_2017}.
Two-dimensional TMDs \cite{wang2012electronics, chhowalla_chemistry_2013} of composition MX$_2$ consist of a triangular lattice of a transition metal M, intercalated between two layers of chalcogen atoms (X = S, Se, Te).
Two high-symmetry configurations of the three atomic planes are possible, leading to a coordination of the transition metal atom exhibiting either trigonal antiprismatic (or distorted octahedral) or trigonal prismatic symmetry.
The two coordinations lead to two families of polymorphs, referred to as $1T$ and $1H$, respectively.

With a few exceptions, all metallic TMDs experience some form of lattice distortion of various strength \cite{manzeli_2d_2017}.
For group V TMDs (M = V, Nb, Ta), characterized by the $d^1$ formal electronic configuration of the transition metal ion~\cite{mattheiss1973band},  the distortions in both polymorphs are weak to moderate, and are usually referred to as charge-density-wave (CDW) phases \cite{wilson_charge-density_1975}. 
On the other hand, the distortions in group VI (M = Mo, W) and VII (M = Tc, Re) TMDs with $d^2$ and $d^3$  formal occupations, in the $1T$ polymorph, are much stronger \cite{wildervanck_dichalcogenides_1971, kertesz1984octahedral}. 

A Peierls mechanism based on the Fermi surface nesting argument \cite{peierls1955quantum, chan1973spin} was originally proposed for $d^1$ TMDs in both polymorphs \cite{wilson_charge-density_1974, wilson_charge-density_1975}, although this point of view has often been challenged in the more recent literature \cite{johannes_fermi_2008}, with several authors arguing that anisotropic momentum-dependant electron-phonon interactions are required to explain the phenomenology \cite{zhu_classification_2015}. 
Real-space chemical bonding arguments have also been proposed \cite{whangbo1992analogies, yi2018coupling}.
Numerous experimental and theoretical studies of CDWs in $d^1$ TMDs have been reported in the last few years, for bulk, few-layers and monolayers forms of these materials \cite{calandra2009effect, ge2012effect, weber2011extended, xi2015strongly, ugeda2016characterization, silva2016electronic, albertini2017effect, battaglia2005fermi, liu2009electron, ge2010first, zhang2014spin, yu2015gate, chen2015influence, shao2016manipulating, yi2018coupling, miller2018charge, sakabe2017direct, kamil2018electronic, calandra2018,  pasquier2018charge, pasztor2017dimensional, zhang2017strain, umemoto_pseudogap_2018, mulazzi2010absence}.
It is striking to note that, while certain authors mention a well-understood nesting mechanism, others consider nesting unimportant \cite{johannes_fermi_2008, liu2009electron, ge2010first, mulazzi2010absence, zhu_classification_2015}.

\begin{figure*}[t]
\centering
\includegraphics[width=16cm]{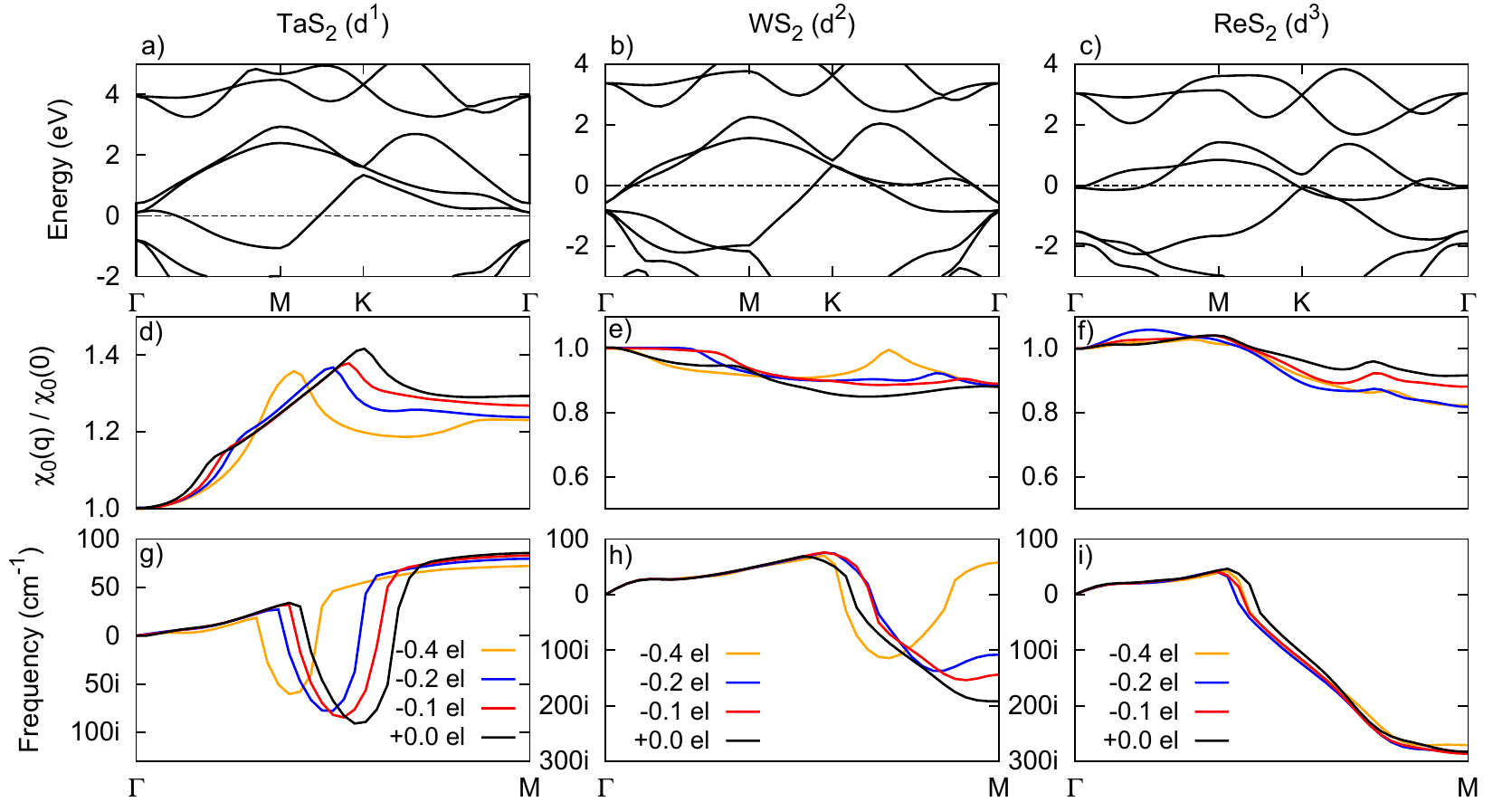}
\caption{\label{fig:bandsphonon} Band structures calculated from first principles for monolayers of (a) $1T$-TaS$_2$, (b) $1T$-WS$_2$, and (c) $1T$-ReS$_2$. The Fermi level is set to zero. Calculated bare static susceptibility along $\Gamma M$ for (d) $1T$-TaS$_2$, (e) $1T$-WS$_2$, and (f) $1T$-ReS$_2$. Calculated dispersion for the lowest acoustic phonon branch along $\Gamma M$ for (g) $1T$-TaS$_2$, (h) $1T$-WS$_2$, and (i) $1T$-ReS$_2$.}
\end{figure*}
Whereas the $1H$ polymorph of $d^2$ TMDs is semiconducting and stable, the $1T$ phase is highly unstable and distorts into the metastable $1T'$ phase, with $2\times 1$ periodicity \cite{whangbo1992analogies, duerloo2014structural}.
The $1T'$ phase of $d^2$ TMDs was recently the focus of intense attention due to its topological properties \cite{qian_quantum_2014, fei2017edge, tang2017quantum, pulkin2017robustness, Ugeda18}, but the mechanism of the distortion has been less discussed. 
A Peierls nesting mechanism was also suggested for certain Mo dichalcogenides \cite{keum_bandgap_2015, shirodkar2014emergence}, based on the inspection of the Fermi surface that reveals pockets apparently nested by the correct wave vectors \footnote{In Ref.~\cite{shirodkar2014emergence}, the calculated instability for MoS$_2$ is maximal at the $K$ point (corresponding to $\sqrt{3}\times\sqrt{3}$ periodicity) instead of the $M$ point. This is due to the use of a too coarse grid of $q$-points for Fourier interpolation. The proposed nesting mechanism in Ref.~\cite{shirodkar2014emergence} is to explain the instability at the $K$ point.}.
TMDs with $d^3$ formal occupation are found in a strongly distorted form of the $1T$ polymorph with $2 \times 2$ periodicity (sometimes referred to as $1T''$), with tetramer clusters of transition metal ions forming diamond chains \cite{wildervanck_dichalcogenides_1971, tongay2014monolayer}. 
Kertesz and Hoffman first derived the structure theoretically and stressed the role of the strong interactions between in-plane $d_{xy}$ and $d_{x^2-y^2}$ electrons in driving the distortion \cite{kertesz1984octahedral}.
In an attempt to provide a unified theory for the distortions in the TMDs, Whangbo and Canadell suggested a complementary picture of both hidden nesting and local chemical bonding \cite{whangbo1992analogies}, as for the $1T'$ phase in $d^2$ TMDs.
More recently, it has been proposed that the $1T''$ phase should be understood as a Peierls instability of the $1T'$ phase, due to the existence in this phase of quasi-1D bands at half-filling for $d^3$ ions \cite{choi2018origin}.

In this Letter, we study, from density functional theory (DFT) calculations, the doping-dependent fermiology and phonon instabilities in $5d$ $1T$ TMDs with increasing $d$-shell population, taking monolayers of the disulfides TaS$_2$, WS$_2$ and ReS$_2$ as examples.
For TaS$_2$, the doping-dependence of the calculated incommensurate CDW (ICDW) wave vector and its correspondence with the bare susceptibility provide a clean demonstration of the effect of the fermiology on the ICDW. 
We therefore argue that at $n\approx1$ $d$ electron ({\it i.e.} TaS$_2$ or heavily hole-doped WS$_2$), a weak-coupling $k$-space nesting picture is still a good starting point for understanding, although no sharp divergence is present in the bare susceptibility.
On the other hand, we show that for $n\approx 2$--$3$ $d$ electrons (WS$_2$ and ReS$_2$), nesting arguments are not useful, and that a real-space strong-coupling picture of bonding Wannier functions (WFs), splitting strongly the $t_{2g}$ triplet, applies and provides a simple physical picture.
This suggests a crossover between weak-coupling and strong-coupling regimes as a function of the electronic filling of the $t_{2g}$ subshell. 

Figs.~\ref{fig:bandsphonon}(a)--\ref{fig:bandsphonon}(c) show the electronic band structures for undistorted monolayers of $1T$-TaS$_2$, $1T$-WS$_2$ and $1T$-ReS$_2$, calculated from first principles in the generalized gradient approximation~\cite{perdew_generalized_1996}.
Details of the first-principles calculations are given in the Supplemental Information \cite{suppl}.
The three bands close to the Fermi level are very similar for the three materials (except for the position of the Fermi level) and have $t_{2g}$ orbital character, i.e. $d_{xy}$, $d_{xz}$ and $d_{yz}$, with the $z$-axis pointing along an M--S bond.
The latter choice of coordinates allows to almost perfectly decouple the two high-energy and three low-energy $d$ orbital degrees of freedom \cite{pasquier2018crystal}, justifying the denomination $t_{2g}^1$ for TaS$_2$, $t_{2g}^2$ for WS$_2$, and $t_{2g}^3$ for ReS$_2$.

Figs.~\ref{fig:bandsphonon}(d)--\ref{fig:bandsphonon}(i) show the calculated bare static susceptibilities and phonon dispersions along the $\Gamma M$ direction, for the three materials and for undoped and hole-doped cases \footnote{For definiteness, we study the hole doping to understand the effect of doping in these materials. The electron-doped case is analogous.}. 
For the sake of clarity, we have only shown the lowest-energy acoustic phonon mode, that softens for the three materials for all doping levels considered.
To evaluate the bare susceptibility, we have adopted the commonly-used constant-matrix-elements approximation (CMA), $\chi_{0}(q) = \frac{1}{N_k}\sum_{k,n,n'}\frac{f_{nk+q}-f_{n'k}}{\epsilon_{nk+q}-\epsilon_{n'k}}$, where $N_k$ is the number of $k$-points in the discretized Brillouin zone, $\epsilon_{nk}$ is the energy of band $n$ at momentum $k$, and $f$ is the Fermi-Dirac distribution.
We have included the three $t_{2g}$-like bands in the summation, and set the electronic temperature to $300$~K. 
Using the CMA, the absolute value of the susceptibility is sensitive to the number of bands included in the summation \cite{heil2014accurate}. 
However, we have verified that the location of the peak for TaS$_2$, as well as the absence of peaks at $M$ for WS$_2$ and ReS$_2$, are robust with respect to the number of bands considered. 

In the theory of weak-coupling charge- and spin-density-wave instabilities, the bare susceptibility is the key quantity. Its enhancement at certain wave vectors favours softening of certain phonon or magnon modes, depending on the dominant microscopic interaction, either electron-phonon or electron-electron \cite{chan1973spin}. 
In the limit of perfect nesting, the bare susceptibility exhibits logarithmic divergences at momentum $2k_F$, leading to instabilities at infinitesimal coupling constant.
In real materials, perfect nesting would require unrealistic fine-tuning, but nesting-derived instabilities can still occur provided the interactions are not too weak.

\begin{figure}[t]
\centering
\includegraphics[width=7cm]{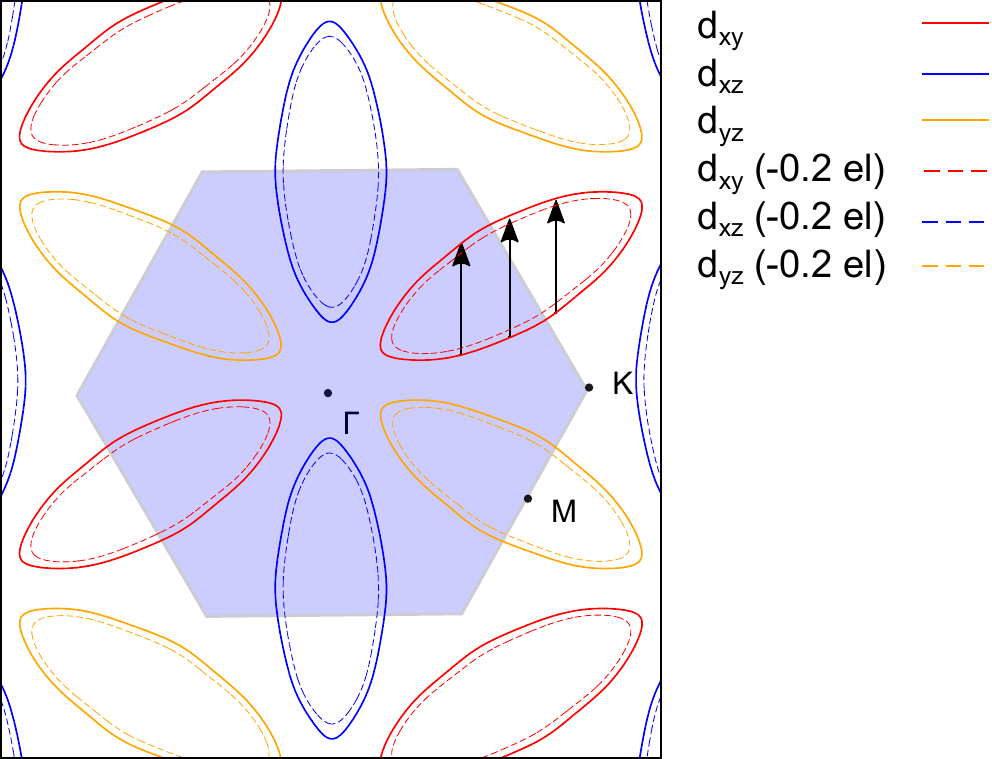}
\caption{\label{fig:tas2} Fermi surface of monolayer $1T$-TaS$_2$ (undoped and hole-doped). The shaded area delimits the Brillouin zone. Nesting vectors for the undoped case have been drawn.}
\end{figure}
Fig.~\ref{fig:bandsphonon}(d) shows that, unlike for most 2D metals, the bare susceptibility of $1T$-TaS$_2$ does not achieve its maximum at the $\Gamma$ point, but at an incommensurate wave vector along the $\Gamma M$ direction, corresponding to the momentum $q_{\mathrm{ICDW}}\approx 0.29 b_i$ (where $b_i$ are the three primitive vector of the reciprocal lattice) where the calculated phonon softening is maximal.
This is due to the approximate nesting properties of the Fermi surface, shown in Fig.~\ref{fig:tas2}.
Moreover, the calculated peak of the susceptibility, as well as the calculated $q_{\mathrm{ICDW}}$, are found sensitive to the exact position of the Fermi level and both change upon doping.
Such behaviour is typical of a $2k_F$ effect and clearly shows the effect of the change of the Fermi surface area upon doping on the ICDW.
Experimentally, Ti-doped bulk $1T$-TaS$_2$ exhibits an ICDW wave vector that decreases with increasing Ti concentration \cite{wilson1974charge, disalvo1975effect, chen2015influence}.  
For 2D materials, electrostatic doping allows inducing charge carriers in a way that closely resembles the rigid Fermi level shift in our calculations.
It would therefore be interesting to address the change of ICDW periodicity in gated TaS$_2$ and other similar materials.  
Bulk TaS$_2$ (and possibly the monolayer as well \cite{albertini2016zone}) undergoes the so-called lock-in transition, where the CDW adopts a periodicity commensurate with the high-symmetry phase, characterized by a commensurate wave vector that corresponds to $\sqrt{13}\times \sqrt{13}$ periodicity
\cite{mcmillan_landau_1975, mcmillan1976theory}.  
We stress that the calculated CDW wave vectors and peaks in the susceptibility correspond to the ICDW periodicity, as the lock-in transition results from anharmonic effects.
 
As Figs.~\ref{fig:bandsphonon}(e)--\ref{fig:bandsphonon}(f) show, the maximum phonon softening for the $t_{2g}^2$ and $t_{2g}^3$ cases occurs at the $M$ point, indicating an instability towards doubling the unit cell.
Compared to TaS$_2$, the phonon softening occurs over a wider range of momenta and is much stronger. 
The phonon softening at the $M$ point is clearly not related to any peak in the bare susceptibility calculated in the CMA. 
Contrary to closely related MoS$_2$ \cite{shirodkar2014emergence} and MoTe$_2$ \cite{keum_bandgap_2015}, the Fermi surface of WS$_2$ does not exhibits nested Fermi pockets, that appear only under electron doping~\cite{suppl} and are therefore not responsible for the instability.
For $n_{t_{2g}}\approx 3$ (ReS$_2$) the phonon instability is robust against doping, so that the calculated soft phonon mode in not sensitive to the exact number of electrons, contrary to the $n_{t_{2g}}\approx 1$ case.
For WS$_2$, the instability at the $M$ point is sensitive to hole doping, and disappears at $n_{\mathrm{hole}}\approx 0.4$.  
For heavily hole-doped WS$_2$, a behaviour analogous to TaS$_2$ is recovered.
Small discommensurations are already present at lower doping, but it is not clear whether these could be observed experimentally because of anharmonic effects.
Clearly, the instability at the $M$ point is not associated with a nesting mechanism, since the calculated susceptibility is at its minimum.
Nesting arguments are perturbative ones, so they become less relevant as the instability grows stronger, as is the case for WS$_2$ and ReS$_2$.

From the considerations above, it appears that lattice distortions in $1T$ $d^2$ and $d^3$ TMDs should be better understood from a strong-coupling perspective.
The strong-coupling qualitative picture of CDWs consists in a real-space picture of chemical bonding \cite{rossnagel2011origin}.
In the following, we shall demonstrate and quantify the bonding mechanism behind the $1T'$ and $1T''$ phases using a Wannier-function approach. 

\begin{figure*}[t]
\centering
\includegraphics[width=17cm]{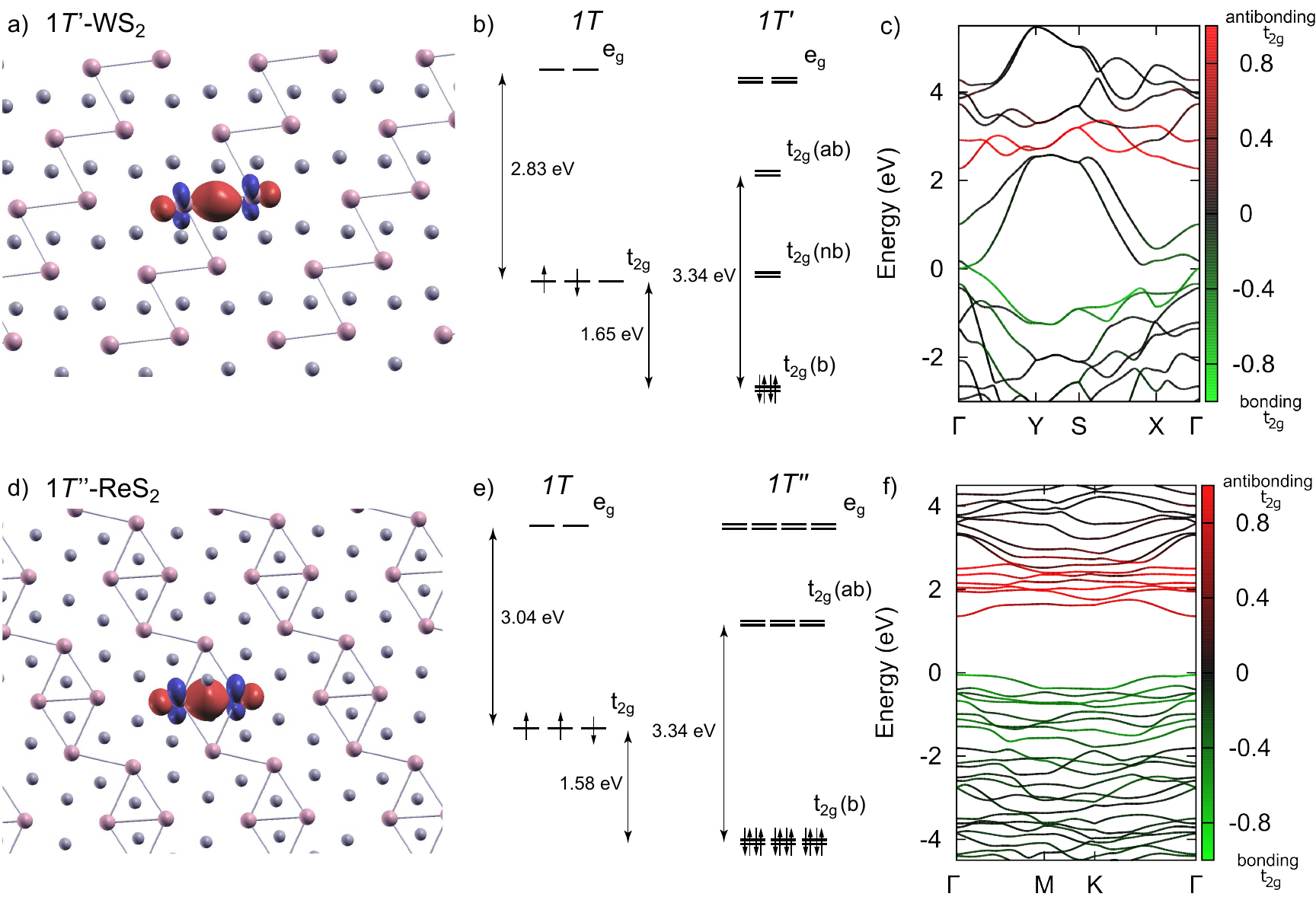}
\caption{\label{fig:ws2} (a) Ball-and-stick representation of the $1T'$ phase of WS$_2$ with an isovalue plot of one of the two equivalent bonding $t_{2g}$ Wannier functions (WFs). W--W bonds have been drawn to facilitate visualization. Each bond accommodates a bonding $t_{2g}$ WF centered on it. (b) Aligned ligand field and modified ligand field energy diagrams for the $1T$ and $1T'$ phases. The bonding (b), nonbonding (nb) and antibonding (ab) $t_{2g}$ states are labeled. (c) Calculated band structure along high-symmetry directions for $1T'$-WS$_2$. The orbital weights of the bonding and antibonding $t_{2g}$ WFs are color-coded. The Fermi level is set to zero. (d)--(f) Corresponding plots for the $1T''$ phase of ReS$_2$.}
\end{figure*}
We begin by discussing the $1T'$ phase of $d^2$ TMDs, taking again WS$_2$ as a representative example.  
The relaxed lattice structure is shown in Fig.~\ref{fig:ws2}(a).
The calculated energy gain upon distortion is large ($0.36$~eV per formula unit), and the change of the electronic structure is drastic.
We have drawn W--W bonds for which the interatomic distance is significantly reduced ($2.78$~\AA\ {\it vs.} $3.21$~\AA\ in the undistorted $1T$ phase).  
Such a large shortening of the W--W distance suggests that $t_{2g}$ states pointing toward these bonds interact strongly with their nearest neighbours, forming bonding and antibonding combinations \cite{whangbo1992analogies}. 
To verify this hypothesis, we construct Maximally Localized Wannier Functions (MLWFs) \cite{marzari_maximally_1997} by considering two different sets of bands separately to assess the formation of bonding states (see Supplemental Information \cite{suppl} for details).

Fig.~\ref{fig:ws2}(b) shows the aligned ligand field (including electrostatic and $pd$ hybridization effects, as we have discussed in Ref.~\cite{pasquier2018crystal}) and modified ligand field energy diagrams for the $1T$ and $1T'$ phases of WS$_2$, obtained using MLWFs \cite{scaramucci2015separating}.
Our Wannier analysis demonstrates that the main effect of the distortion is to split strongly the $t_{2g}$ states into bonding, nonbonding and antibonding WFs, while the $e_g$ states are weakly affected, although the lifting of degeneracy within the $e_g$ doublet is somewhat increased ($0.36$~eV {\it vs.} $0.05$~eV in the $1T$ phase).
In Fig.~\ref{fig:ws2}(a), we show an isovalue plot of one of the two equivalent bonding $t_{2g}$ WFs, centered on a W--W bond (other WFs plots are presented in the Supplemental Material \cite{suppl}).
The on-site energies of the nonbonding $t_{2g}$ states, pointing in the direction of the zigzag chain, are found to be very close ($\sim$0.1~eV difference) to these of the undistorted $1T$ phase.
On the other hand, the $t_{2g}$ WFs pointing in the W--W bonds directions are split in energy by $3.34$~eV. 
The calculated energy splitting is significantly larger than the half-bandwidth of the undistorted $1T$ phase ($W/2 \approx 2.23$~eV), that one would obtain by simply doubling the unit cell without distortion. 
This indicates the formation of strong W--W bonds upon translational symmetry breaking. 
Moreover, Fig.~\ref{fig:ws2}(c) shows that the two bonding $t_{2g}$ WFs contribute mainly to the two occupied bands closest to the Fermi level, and are therefore roughly filled by two electrons. 
The optimal filling of the two strongly bonding WFs explains why the $1T'$ phase is energetically favourable for $n_{t_{2g}}\approx 2$. 

Let us now consider the diamond-chain structure (or the $1T''$ phase) of $d^3$ $1T$ TMDs with $2\times 2$ periodicity, with ReS$_2$ taken as an example. 
The relaxed structure in the $2\times2$ supercell, shown in Fig.~\ref{fig:ws2}(d), is associated with a large energy gain of $1.12$~eV/f.u. compared to the undistorted $1T$ phase.
We have drawn Re--Re bonds, because the interatomic distance between the corresponding Re atoms is significantly reduced compared to the undistorted phase (2.71--2.9~\AA\ {\it vs.} 3.1~\AA\ in the $1T$ phase).  

As for WS$_2$, we have constructed MLWFs by considering separately two sets of bands \cite{suppl}. 
The aligned ligand field and modified ligand field energy diagrams for the $1T$ and $1T''$ phases are represented in Fig.~\ref{fig:ws2}(e).
The whole $t_{2g}$ subshell is strongly split into bonding and antibonding states in the $1T''$ phase.
Indeed, we estimate an energy splitting of $3.34$~eV, significantly larger than the half-bandwidth of the undistorted $1T$ phase ($W/2 \approx 2.22$~eV).
Since not all the shortened bonds are equal in the $1T''$ phases, there are differences in the on-site energies of the corresponding WFs. 
The bonding WF on the shortest bond ($2.71$~\AA), plotted in Fig.~\ref{fig:ws2}(d), is found $0.24$~eV lower in energy compared to that centered on the longest bond ($2.9$~\AA).
As Fig.~\ref{fig:ws2}(f) shows, the bonding $t_{2g}$ WFs contribute mostly to the top of the occupied-bands manifold.
Hence, in the $1T''$ phase at $t_{2g}^3$, all the strongly bonding $t_{2g}$ WFs are fully occupied, explaining the stability of this phase. 

In summary, we report a first-principles study of doping-dependent fermiology and phonon instabilities in 2D $1T$ transition metal disulfides at $d^1$, $d^2$, and $d^3$ occupation of the $d$ shell.
When the electron filling of the $t_{2g}$ subshell is well below half-filling, as in TaS$_2$, we find that the dependence of the ICDW wave vector on the doping levels matches that of the peak of the bare susceptibility.
This behaviour is suggestive of a $2k_F$ effect and supports the view that a $k$-space nesting picture is a good, and necessary, starting point for understanding, even though this point of view has often been challenged.
When the electron filling of the $t_{2g}$ subshell is closer to half-filling, as in WS$_2$ and ReS$_2$, the behaviour is qualitatively different and nesting appears irrelevant.
Our Wannier-function analysis shows that the effect of the distortions is mainly to split strongly the $t_{2g}$ states, and that simple bond-counting arguments are qualitatively correct.
Our study thus provides a unifying picture of lattice distortions in $1T$ TMDs that bridges two regimes, while the crossover between these regimes can be attained by tuning the electron filling of the $t_{2g}$ orbitals. 
Although our study considers monolayer transition metal disulfides as examples, the universality of the electronic structure of TMDs allows to extend our reasoning  to other member of this family of materials, with certain ditellurides as possible exceptions, and to bulk and multilayer materials owing to relatively weak interlayer coupling. \\
 
\noindent
We acknowledge funding by the European Commission under the Graphene Flagship (grant agreement No.~696656). 
We thank QuanSheng Wu for technical assistance.
First-principles calculations were performed at the facilities of Scientific IT and Application Support Center of EPFL.
\bibliographystyle{apsrev4-1}
\bibliography{manuscript}
\end{document}